\documentclass[aps,amsmath,nofootinbib,amssymb,superscriptaddress,twocolumn,prl]{revtex4}
\usepackage{amssymb,graphics,graphicx,color,epstopdf,subfigure,dcolumn,bm,slashed,amsmath}

\begin{document}
{\hspace{+8cm} KEK-TH-2058\quad PITT-PACC-1811\quad  OU-HET-974}
\title{ Probing the Higgs Yukawa coupling to the top-quark at the LHC \\via single top + Higgs production}
\author{Vernon Barger}\email {barger@pheno.wisc.edu}
\affiliation{Department of Physics, University of Wisconsin, Madison, WI 53706, USA}
\author{Kaoru Hagiwara} \email{kaoru.hagiwara@kek.jp}
\affiliation{KEK Theory Center and Sokendai, Tsukuba, Ibaraki 305-0801, Japan}
\affiliation{PITT-PACC, Department of Physics and Astronomy, University of Pittsburgh, PA 15260, USA}
\author{Ya-Juan Zheng} \email{yjzheng@het.phys.sci.osaka-u.ac.jp}
\affiliation{Department of Physics, Osaka University, Osaka 560-0043, Japan}

\begin{abstract}
The conjoined production at the LHC of single top and Higgs boson via $t$-channel weak boson exchange is ideal to probe the top-quark Yukawa coupling, due to a delicate cancellation between the amplitudes with the $htt$ and the $hWW$ couplings. We find that the top quark is produced with $100\%$ polarization in the leading order, and its quantum state is determined by the spin-vector direction in the $t$-quark rest frame. We relate the spin direction to the four-momenta of the top, Higgs and a jet in the helicity amplitude framework. We identify a polarization asymmetry that is sensitive to CP violation, even after partial integration over the forward jet momentum.
This CP violating asymmetry may be observed at the LHC via the component of the top-quark polarization that is perpendicular to the $th$ scattering plane.
\end{abstract}
\maketitle

The coupling of the 125 GeV Higgs boson ($h$) to the top quark, which is the largest of the Standard Model (SM) couplings, is an important target of the LHC experiments. Measurements of the loop-induced $hgg$ and $h\gamma\gamma$ transitions constrain $htt$ indirectly, but these are subject to possible contributions from new physics loops beyond the SM.
 Direct measurements of $htt$ at the LHC can be made through the QCD production of a $t\bar{t}$ pair and $h$, and also through the electroweak production of single $t$ (or $\bar{t}$) and $h$. 
 The latter process proceeds via $t$-channel $W$ exchange. It is particularly promising, because the SM production cross section in $pp$ collisions at 13 TeV c.m.\ energy are sizeable, 48.8 fb in Next to Leading Order (NLO) for $t+h$ and 25.7 fb for $\bar{t}+h$~\cite{Demartin:2015uha}, and also because
 the prediction is known to be extremely sensitive to the relative sign of the $htt$ and $hWW$ couplings~\cite{Stirling:1992fx,Bordes:1992jy}.
The total $t+h$ cross section becomes more than 10 times larger than the SM prediction if the sign of the $htt$ coupling is reversed from the SM value. 
This extreme sensitivity is due to the cancellation between the amplitudes with the $htt$ coupling and those with the $hWW$ coupling, which thereby enhances a probe of a non-SM $htt$ coupling through the interferences with the amplitudes of the well constrained $hWW$ coupling. 
Tentative attempts in understanding this structure for single top plus Higgs production at hadron colliders and the QCD background~\cite{Maltoni:2001hu}, with Higgs decay channels $h\to WW/ZZ$~\cite{Barger:2009ky}, $\gamma\gamma$~\cite{Biswas:2012bd,Yue:2014tya,Gritsan:2016hjl}, and $ b\bar{b}$~\cite{Farina:2012xp} have been performed. CP phases of top Yukawa coupling are studied in~\cite{Biswas:2012bd,Yue:2014tya,Gritsan:2016hjl,Rindani:2016scj}. 

\begin{figure}[t]
\begin{centering}
\begin{tabular}{c}
\includegraphics[width=0.25\textwidth]{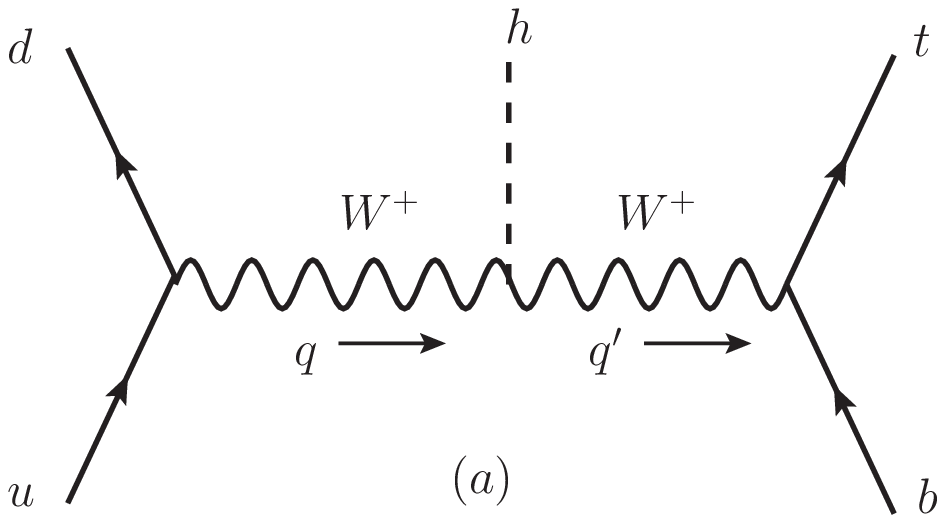}\label{fig:1a}
\includegraphics[width=0.25\textwidth]{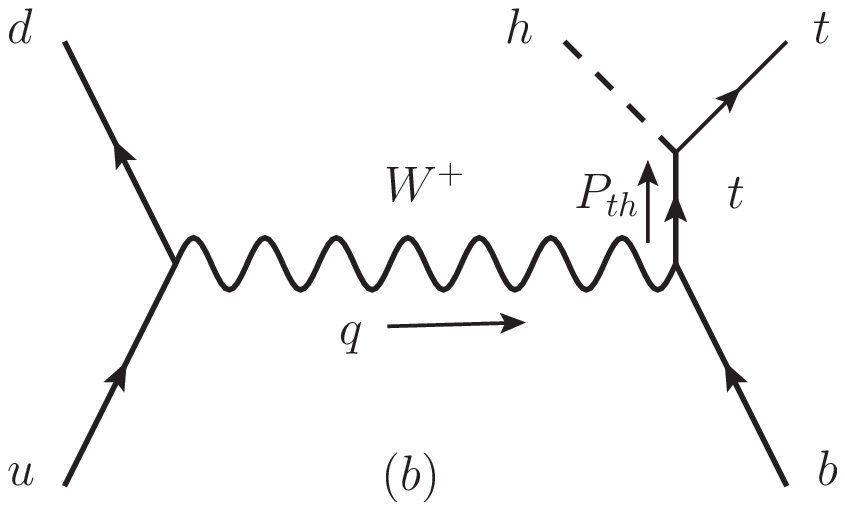}\label{fig:1b}
\end{tabular}
\caption {
Feynman diagrams of $ub\to dth$ process. The four momenta $q^\mu$ and $q^{\prime \mu}$ along the $W^+$ and $P_{th}^\mu$ along the top propogators are shown with arrows.}
\label{fig:feyn}
\end{centering}
\end{figure}

In this letter, we present the helicity amplitudes of the processes
\begin{eqnarray}
ub\to dth\quad\text{and}\quad\bar{d}b\to\bar{u}th \label{eq:1}
\end{eqnarray}
in the massless $b$-quark approximation, from which we can obtain all possible observables that can probe the Higgs couplings. 

We adopt the following minimal non-SM modification to the top Yukawa coupling,
\begin{eqnarray}
{\cal L}&=&-g_{htt}h\bar{t}(\cos\xi_{htt}+i\sin\xi_{htt}\gamma_5)t
\label{eq:lag}
\end{eqnarray}
where $g_{htt}=(m_t/v)\kappa_{htt}$, $(\kappa_{htt}>0)$ and $-\pi\leq\xi_{htt}\leq\pi$. 
The SM values are $\kappa_{htt}=1$ and $\xi_{htt}=0$. 
 CP invariance is violated in Eq.\ (\ref{eq:lag}) when $\sin\xi_{htt}\neq0$, so we study observables that are proportional to $\sin\xi_{htt}$, as signals of CP violation. 

The Feynman diagrams of the sub-process $ub\to dth$ are shown in Fig.~\ref{fig:feyn}. The left diagram~($a$) has the $hWW$ coupling, while the right diagram ($b$) has the $htt$ coupling. The $u\to dW^+$ emission part is common to both diagrams. By combining the $u\to dW^+$ emission amplitudes with the $W^+b\to th$ amplitudes in the $W^+b$ rest frame, we obtain the full helicity amplitudes for the process $ub\to dth$\footnote{The amplitudes (\ref{eq:amp}) agree exactly with the numerical HELAS code~\cite{Murayama:1992gi} which is generated by Madgraph~\cite{Alwall:2014hca}.}. 

\begin{subequations}\label{eq:amp}
{\allowdisplaybreaks
\begin{align}
M_+^{}&=\frac{1-\tilde{c}}{2}e^{i\phi}\sin\frac{\theta^\ast}{2}A\frac{1+\cos\theta^\ast}{2}\nonumber\\
&+\frac{1+\tilde{c}}{2}e^{-i\phi}\sin\frac{\theta^\ast}{2}\Bigg[A\left(\frac{1+\cos\theta^\ast}{2}+\epsilon_1\right)\nonumber\\
&\quad\quad\quad\quad\quad\quad\quad\quad
-B\left(e^{-i\xi}+\delta\delta^\prime e^{i\xi}\right)\Bigg]\nonumber\\
&+\frac{\tilde{s}}{2}\cos\frac{\theta^\ast}{2}\frac{W}{Q}
\Bigg[A\left(\frac{q^\ast E_h^\ast+q^{0\ast}p^\ast\cos\theta^\ast}{Wp^\ast}+\epsilon_1\right)
\nonumber\\
&\quad\quad\quad\quad\quad\quad\quad\quad
-B\left(e^{-i\xi}+\delta\delta^\prime e^{i\xi}\right)\Bigg],
\label{eq:M+}\\
M_-^{}&=-\frac{1-\tilde{c}}{2}e^{i\phi}\cos\frac{\theta^\ast}{2}A\delta\frac{1-\cos\theta^\ast}{2}\nonumber\\
&-\frac{1+\tilde{c}}{2}e^{-i\phi}\cos\frac{\theta^\ast}{2}
\Bigg[A\left(\delta\frac{1-\cos\theta^\ast}{2}-\epsilon_2\right)
\nonumber\\
&\quad\quad\quad\quad\quad\quad\quad\quad
+B\left(\delta e^{-i\xi}+\delta^\prime e^{i\xi}\right )\Bigg]\nonumber\\
&-\frac{\tilde{s}}{2}\sin\frac{\theta^\ast}{2}\frac{W}{Q}
\Bigg[A\left(\delta\frac{q^\ast E_h^\ast+q^{0\ast}p^\ast\cos\theta^\ast}{Wp^\ast}+\epsilon_2\right)
\nonumber\\
&\quad\quad\quad\quad\quad\quad\quad\quad
-B\left(\delta e^{-i\xi}+\delta^\prime e^{i\xi}\right)\Bigg].
\label{eq:M-}
\end{align}}
\end{subequations}
In the Breit frame~\cite{Hagiwara:2009wt}, the $u$ and the $d$ quark four momenta are specified by
\begin{subequations}
\begin{align}
p_u^\mu&=\tilde{\omega}(1,\sin\tilde{\theta}\cos\phi,-\sin\tilde{\theta}\sin\phi,\cos\tilde{\theta}),\\
p_d^\mu&=\tilde{\omega}(1,\sin\tilde{\theta}\cos\phi,-\sin\tilde{\theta}\sin\phi,-\cos\tilde{\theta}),
\label{eq:pud}
\end{align}
\end{subequations}
 where $2\tilde{\omega}\cos\tilde{\theta}=Q$ and $2\tilde{\omega}=Q\left(2\hat{s}/(W^2+Q^2)-1\right)$, 
 $\hat{s}=(p_u+p_b)^2$ and $W=\sqrt{P_{th}^2}=\sqrt{(p_t+p_h)^2}$. 
 The factors $A$ and $B$ normalize the $hWW$ and $htt$ contributions, respectively,
\begin{subequations}\label{eq:AB}
\begin{align}
A&=2g^2D_W^{}(q)\tilde{\omega}\sqrt{2q^\ast(E^\ast+p^\ast)}\frac{mp^\ast}{m_W^2}g_{hWW}^{}D_W(q^\prime)\label{eq:A},\\
B&=-2g^2D_W^{}(q)\tilde{\omega}\sqrt{2q^\ast (E^\ast+p^\ast)}Wg_{htt}^{}D_t^{}(P_{th})\label{eq:B}.
\end{align}
\end{subequations}
 We introduce the azimuthal angle about the common $\vec{q}$ axis between the $u\to dW^+$ emission plane and the $W^+b\to th$ production plane. With the orientation of Eq.~(\ref{eq:pud}), the $t$ momentum is in the $z$-$x$ plane with $\phi=0$. 

 The negative sign in $B$ makes both $A$ and $B$ positive, since the propagator factors $D_W^{}(q)=1/(q^2-m_W^2)$ and $D_W(q^\prime)$ are negative with $q^\prime=q-p_h$, while $D_t^{}(P_{th})=1/(P_{th}^2-m_t^2)$ is positive. We keep the $hWW$ coupling $g_{hWW}=(2m_W^2/v)\kappa_{hWW}$ standard ($\kappa_{hWW}=1$) in this report. We introduce notation for $\tilde{c}=\cos\tilde{\theta}$ and $\tilde{s}=\sin\tilde{\theta}$ for the Breit frame angles, while $\xi=\xi_{htt}$. The starred momenta are defined in the $th$ rest frame,
 $q^\mu=(q^{0\ast},0,0,q^\ast),~
p_t^\mu=(E^\ast,p^\ast\sin\theta^\ast,0,p^\ast\cos\theta^\ast)$
and $E_t^\ast+E_h^\ast=q^{0\ast}+q^\ast=W$ gives the invariant mass of the $th$ system. 
The factors $\delta=m_t/(E^\ast+p^\ast)$ and $\delta^\prime=m_t/W$,
$\epsilon_1=m_W^2/[p^\ast(E^\ast+p^\ast)],\quad \text{and}~\epsilon_2=m_W^2/(m_tp^\ast)$, are small at large $W$.

All the $\theta^\ast$ dependences of the amplitudes, except for those in $D_W(q^\prime)$, are expressed in terms of $J=1/2$ and 3/2 $d$ functions. In particular, the first term in Eq.~(\ref{eq:M+}) and Eq.~(\ref{eq:M-}) give amplitudes for the collision of $\lambda=+1$ $W^+$ and the helicity $-1/2$ $b$ quark, and hence only $J_z^{}=3/2$ $d$ functions appear, with no $s$-channel top contribution. The second ($\lambda=-1$) and the third ($\lambda=0$) terms have both $t$-channel $W$ and $s$-channel top propagator amplitudes. More importantly, we note the $\lambda=0$ (longitudinal $W$) enhancement factor of $W/Q$ in both amplitudes. It's typical value is $W/Q\sim6$ since the cross section at $\sqrt{s}=13$ TeV peaks at $W\sim 350$ GeV and $Q\sim60$ GeV when $p_T^{}>30$ GeV forward jet tag is applied. In the high $W$ limit where $\delta=\delta^\prime$, the amplitudes are proportional to the factors 
\begin{subequations}\label{eq:simAB}
\begin{align}
M_+&\sim\frac{W}{2Q}\sin\tilde{\theta}\cos\frac{\theta^\ast}{2}\left[\frac{1+\cos\theta^\ast}{2}A-e^{-i\xi}B\right]
,\\
M_-&\sim-\frac{W}{2Q}\sin\tilde{\theta}\sin\frac{\theta^\ast}{2}\left[\frac{1+\cos\theta^\ast}{2}A-2\cos\xi B\right]\delta,
\end{align}
\end{subequations}
where we keep the relative phase and the normalizations of the amplitudes. The helicity $+1/2$ top amplitudes $M_+$ dominate at high $W$ because of the chirality flip Yukawa coupling from the left handed $b$ quark, including the Goldstone component of the second $W^+$ propagator in the diagram Fig.~(1a), while $M_-$ is suppressed by the top helicity flip factor, $\delta=m_t/(E^\ast+p^\ast)$. The destructive interference for $\xi=0$ is manifest in both amplitudes because both $A$ and $B$ in Eqs.\ (\ref{eq:AB}) are positive definite. We further note that the amplitude $M_-$ is almost real because $\delta e^{-i\xi}+\delta^\prime e^{i\xi}\sim 2\delta\cos\xi$ at large $W$, while $M_+$ can become complex, being proportional to $e^{-i\xi}=\cos\xi-i\sin\xi$ in the same limit.

We note in passing that the amplitudes for the process $cb\to sth$ are exactly the same as those of $ub\to dth$ in Eq.~(\ref{eq:amp}), whereas those for the process $\bar{d}b\to\bar{u}th$ and $\bar{s}b\to \bar{c}th$ are obtained from Eq.~(\ref{eq:amp}) simply by changing the Breit frame angle, $\cos\tilde{\theta}\to-\cos\tilde{\theta},~(\tilde{c}\to-\tilde{c})$.
This does not affect the leading part of the $\lambda=0$ helicity amplitude, but it changes the subleading transverse $W$ amplitude such that $e^{-i\phi}$ becomes $e^{i\phi}$. Therefore antiquark contribution to the $th+j$ process reduces the asymmetry in $\phi$ distributions.

\begin{figure}[t]
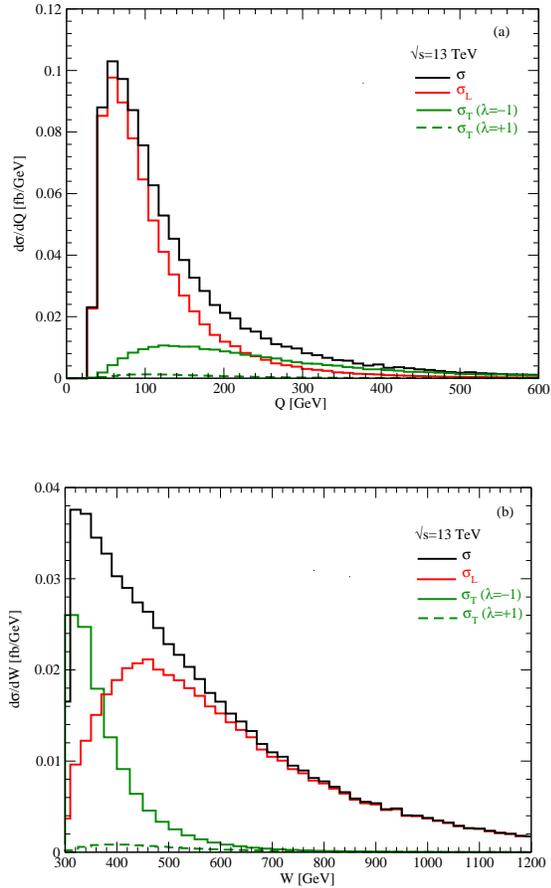

\begin{centering}
\begin{tabular}{c}
\vspace{+0.8cm}
\includegraphics[width=0.4\textwidth]{Q.eps}\\
\includegraphics[width=0.4\textwidth]{W.eps}
\end{tabular}
\caption {$d\sigma/dQ$ (upper) and $d\sigma/dW$ (lower). $Q=\sqrt{-q^2}$ is the invariant momentum transfer of the virtual $W^+$, $W=\sqrt{P_{th}^2}=m(th)$ is  the invariant mass of $th$ system. The red curves show contributions of the longitudinal $W(\lambda=0)$, while the green curves show those of the transverse $W(\lambda=\pm1)$.}
\label{fig:QW}
\end{centering}
\end{figure}

Let us now study the property of the amplitudes quantitatively. Fig.~\ref{fig:QW} shows $d\sigma/dQ~(a)$ and $d\sigma/dW~(b)$ of the subprocess $ub\to dth$, convoluted with the $u$ and $b$ PDF~\cite{Dulat:2015mca} in $pp$ collisions at $\sqrt{s}=13 $ TeV. We set the factorization scale at $\mu=(m_t+m_h)/4$ and impose cuts on $d$ jet at $p_T^j > 30$ GeV, $|\eta_j|<4.5$ to reproduce the results of Ref.~\cite{Demartin:2015uha} in the LO. Shown by red and green curves are the contribution of the longitudinal ($\lambda=0$) and the transverse ($\lambda=\pm1$) $W$ contributions. It is clearly seen that $W_L^{}$ dominates at low $Q$ ($Q\lesssim 100$ GeV) and large $W~(W\gtrsim 400$ GeV), while $W_T(\lambda=-1)$ contribution is significant at large $Q$ ($Q>100$ GeV) and small $W$ ($W<400$ GeV), as expected from our analytic amplitudes.

\begin{figure}[t]
\begin{centering}
\begin{tabular}{c}
\includegraphics[width=0.4\textwidth]{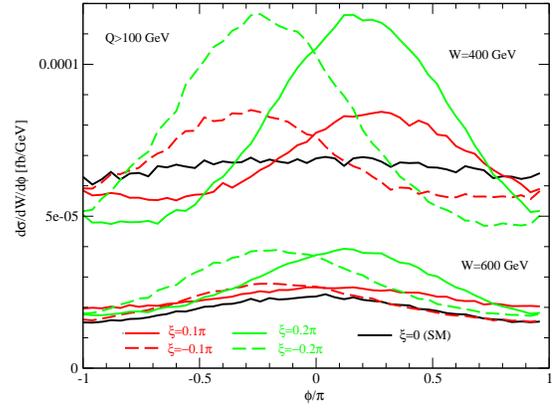}\label{fig:phi2}
\end{tabular}
\caption {$d\sigma/dW/d\phi$ v.s.\ $\phi$ at $W=400$ and 600 GeV for $Q>100$ GeV. Black, red and green curves are for the SM ($\xi=0$), $\xi=\pm0.1\pi,~{\rm and}~\pm0.2\pi$.}
\label{fig:phi}
\end{centering}
\end{figure}

In Fig.~\ref{fig:phi}, we show distributions of the azimuthal angle between the $u\to dW^+$ emission plane and the $W^+b\to th$ production plane about the common $W^+$ momentum direction in the $W^+ b$ rest frame. The results are shown at $W=400$ and 600 GeV for large $Q$ ($Q >$ 100 GeV). The black, red and green curves are for the SM ($\xi=0$), $\xi=\pm0.1\pi,~{\rm and}~\pm0.2\pi$, respectively. Solid curves are for $\xi\geq0$ while dashed curves are for $\xi<0$. 

The $\phi$ distributions are proportional to 
\begin{eqnarray}
|M_+|^2+|M_-|^2
\end{eqnarray} 
where the top polarization is summed over. The interference between the $\lambda=0$ and $\lambda=-1$ amplitudes gives terms proportional to $\sin\phi\sin\xi$, leading to the asymmetry
\begin{eqnarray}\label{eq:asy}
\int^\pi_0d\phi\frac{d\sigma}{d\phi}-\int^{0}_{-\pi}d\phi\frac{d\sigma}{d\phi}
\end{eqnarray}
that determines the sign of $\sin\xi$\footnote{Asymetries propotional to $\sin\xi$ can be regarded as indicators of CP violation in the process, whereas the $|\xi|$ dependences in total and differential cross sections can be mimicked e.g.\ by higher dimensional operators.}. The asymmetry is large at small $W$ and large $Q$ because the subleading $\lambda=-1$ amplitudes are significant there, see Fig.\ \ref{fig:QW}. The asymmetry remains significant at $W=400$ GeV, however, even for events with $Q<100$ GeV~\cite{VBKHYZ}.

We are now ready to discuss the polarization of the top quark in the single top$+h$ production processes. We first note that the helicity amplitudes $M_+$ and $M_-$ in Eq.~(\ref{eq:amp}) are purely complex numbers when production kinematics ($\sqrt{s},~Q,~W,\cos\tilde{\theta},~\cos\theta^\ast,\phi$) are fixed. This is a peculiar feature of the SM where only the left-handed $u,~d,~{\rm and}~b$ quarks contribute to the process. It implies that the produced top quark polarization state is expressed as the superposition
\begin{eqnarray}
\left|t\right\rangle=\frac{M_+^{}\left|J_z=+\frac{1}{2}\right\rangle+M_-^{}\left|J_z=-\frac{1}{2}\right\rangle}{\sqrt{|M_+^{}|^2+|M_-^{}|^2}}
\end{eqnarray}
in the top quark rest frame, where the quantization axis is along the top momentum direction in the $th$ rest frame. The top quark is hence in the pure quantum state with $100\%$ polarization, with its orientation fixed by the complex number $M_-/M_+$. Its magnitude determines the polar angle and $arg(M_-/M_+)$ determines the azimuthal angle of the top spin direction.  Therefore, the kinematics dependence of the polarization direction can be exploited to measure the CP phase $\xi$, e.g.\ by combining the matrix element methods with the polarized top decay density matrix\footnote{The top quark decay polarization density matrices for its semi-leptonic and hadronic decays are given e.g.\ in Appendix A of Ref.~\cite{Hagiwara:2017ban}.}.

In this letter, we investigate the prospects of studying CP violation in the $htt$ coupling through the top quark polarization asymmetry in the single $t+h$ process, with partial integration over the final state phase space. 

For this purpose, we introduce a matrix distribution
\begin{eqnarray}
d\sigma_{\lambda\lambda^\prime}^{}=\int dx_1 dx_2 D_{u/p}(x_1) D_{b/p}(x_2) \frac{1}{2\hat{s}} \overline{\sum} M_\lambda M_{\lambda^\prime}^\ast d\Phi_{dth}\nonumber\\
\end{eqnarray}
where the energy fractions $(x_1,x_2)$ and 3-body phase space $d\Phi_{dth}$ can be constrained to give kinematical distributions,  $d\sigma=d\sigma_{++}+d\sigma_{--}$. The polarization density matrix is 
\begin{eqnarray}
\rho_{\lambda\lambda^\prime}^{}
=\frac{d\sigma_{\lambda\lambda^\prime}^{}}{d\sigma_{++}+d\sigma_{--}}
=\frac{1}{2}\left[\delta_{\lambda\lambda^\prime}+\sum_{k=1}^{3}P_k\sigma_{\lambda\lambda^\prime}^k\right]
\end{eqnarray}
for an arbitrary distribution. The coefficients of the three sigma matrices makes a three-vector, $\vec{P}=(P_1,P_2,P_3)$, whose magnitude $P=\sqrt{\vec{P}\cdot \vec{P}}$ gives the degree of polarization ($P=1$ for $100\%$ polarization, $P=0$ for no polarization), while its spatial orientation gives the direction of the top quark spin in the top rest frame. For the helicity amplitudes (\ref{eq:amp}) calculated in the $th$ rest frame, the $z$-axis is along the top momentum in the $th$ rest frame, and the $y$-axis is along the $\vec{q}\times\vec{p}_t$ direction, perpendicular to the $W^+ b\to th$ scattering plane.

\begin{figure}[t]
\begin{centering}
\begin{tabular}{c}
\includegraphics[width=0.4\textwidth]{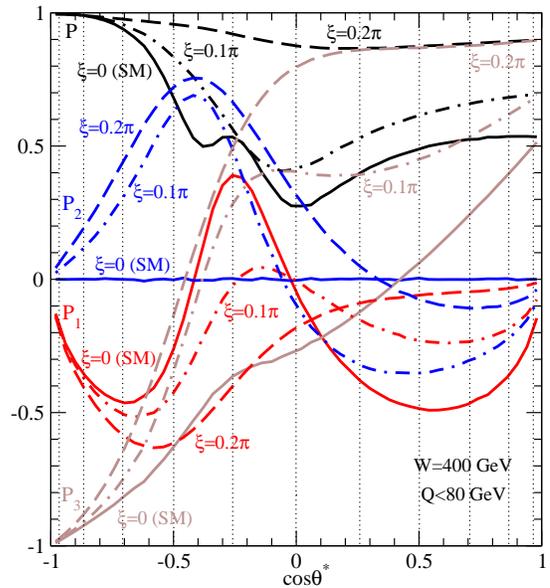}
\end{tabular}
\caption {Predicted top-quark polarization parameters $ P_1$,~$P_2$,~$P_3$ and $P=\sqrt{P_1^2+P_2^2+P_3^2}$ v.s.\ $\cos\theta^\ast$ in the $W^+b\to th$ scattering plane at $W=400$ GeV for $Q\lesssim m_W^{}$. $P_2$ (denoted by the blue curves) is the polarization component perpendicular to the scattering plane. $P_2$ is non-zero if CP is violated.}
\label{fig:pol4}
\end{centering}
\end{figure}

We show in Fig.~\ref{fig:pol4} the degree of polarization $P$ and its three components ($P_1,~P_2,~P_3$) at $W=400$ GeV versus the top scattering angle $\cos\theta^\ast$, when all the other kinematical variables are integrated over subject to the constraint $Q<80$ GeV. The $Q<80$ GeV restriction makes $\lambda=0$ (longitudinal $W$) components dominate the amplitudes. Since the integration over the azimuthal angle $\phi$ kills interference between different $\lambda$ amplitudes, the polarization given in Fig.~\ref{fig:pol4} shows essentially the interference of the $\lambda=0$ components in $M_+$ and $M_-$. At $\cos\theta^\ast=-1 ~(\theta^\ast=\pi)$, $M_-$ dominates over $M_+$ because $\sin\frac{\theta^\ast}{2}=1$ and $\cos\frac{\theta^\ast}{2}=0$ in Eq.~(\ref{eq:amp}). $P_3=-1$, and hence $P=1$. $M_+$ grows quickly as $\cos\theta^\ast$ deviates from $-1$, and the interference between $M_+$ and $M_-$ gives non-trivial polarization of the top quark. Most notably, $P_2=0$ for the SM ($\xi=0$). The top quark polarization lies in the scattering plane when no phase appears in the amplitudes. Strikingly, the polarization perpendicular to the scattering plane, $P_2$ grows quickly as $\xi$ becomes non-zero. Fig.~\ref{fig:pol4} shows that $P_2$ reaches 0.7 at $W=400$ GeV for $\xi=0.1\pi$. If $\xi=-0.1\pi$, instead, the sign of $P_2$ is reversed. The differential cross section is large near $\cos\theta^\ast=-1$, because of the $u$-channel $W$ propagator factor $D_W(q^\prime)$ in $A$ in Eq.\ (\ref{eq:A}). $P_2$ is uniformly positive in the region $\cos\theta^\ast\lesssim0$ with $W\gtrsim 400$ GeV and $Q\lesssim100$ GeV~\cite{VBKHYZ}. It should also be noted that the $W_L^{}$ dominance at low $Q$ region is amplified with $\xi\neq0$, because the destructive interference between the $A$ and $B$ terms in Eq.~(\ref{eq:simAB}) weakens. Accordingly, the degree of polarization $P$ exceeds $90\%$ over the entire $\cos\theta^\ast$ region for $|\xi|\gtrsim0.2\pi$ at $W=400$ GeV.

We therefore propose that the top quark polarization component perpendicular to the scattering plane be measured subject to the restriction $Q\lesssim m_W^{}$.

There is a notable advantage in a $P_2$ measurement in $pp$ collisions, in that this allows CP violation to be clearly disentangled from $T$-odd ($T_N$-odd) asymmetries. The azimuthal angle asymmetry in Eq.~(\ref{eq:asy}) may be regarded as expectation value of the $T_N^{}$-odd product
\begin{eqnarray}
\vec{p}_u\times\vec{p}_d\cdot\vec{p}_t
\end{eqnarray}
where the product $\vec{p}_u\times\vec{p}_d$ determines the $u\to dW^+$ emission plane with orientation. Likewise, the polarization asymmetry $P_2$ is proportional to 
\begin{eqnarray}
\vec{q}\times\vec{p}_t\cdot\vec{s}_t
\end{eqnarray}
where the product $\vec{q}\times\vec{p}_t$ defines the $W^+b\to th$ scattering plane with orientation. Both asymmetries are $T_N^{}$-odd, and hence receive contributions from the final state interaction phases.
Because the processes~(\ref{eq:1}) have color singlet $W$ exchange, the QCD rescattering phase appears only at the two-loop level. The electroweak phase appears in the one-loop level and part of it can be approximated by the width of the $s$-channel top propagator, $D_t(P_{th})=1/(P_{th}^2-m_t^2+im_t\Gamma_t)$. Although we can calculate the SM contributions to the above $T_N$-odd asymmetries, we can disentangle the absorption and CP phases contributions experimentally by measuring $P_2$ for both $th$ production and $\bar{t}h$ production. The key observation is that the asymmetry $P_2$ given in Fig.~\ref{fig:pol4} is essentially the asymmetry of the process 
\begin{eqnarray}\label{eq:W+}
W_L^+ b\to th
\end{eqnarray}
where $W_L^{}$ stands for the $\lambda=0$ component, whereas the asymmetry of the $\bar{t}h$ process is governed by 
\begin{eqnarray}\label{eq:W-}
W_L^- \bar{b}\to \bar{t}h
\end{eqnarray}
Because the processes~(\ref{eq:W+}) and~(\ref{eq:W-}) are CP conjugates, the CP-phase ($\xi$) contribution to $P_2$ are opposite.  The difference between the $P_2$ values gives CP violation, since rescattering contributions cancel.  
This gives a rare opportunity for direct measurement of CP violation in $pp$ collisions. 
\begin{figure}[t]
\begin{centering}
\begin{tabular}{c}
\includegraphics[width=0.45\textwidth]{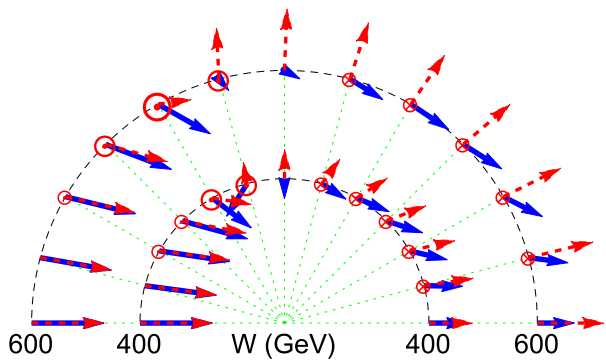}
\end{tabular}
\end{centering}
\caption{Flow map of the top-quark polarization in the $W^+b\to th$ scattering plane. Here, the $th$ c.m.\ energy, denoted by $W$, specifies the radius of the semi-circle, and the polarizations are shown at increments in polar angle $\theta^\ast=n\pi/12~(n=0~\text{to}~12)$. The arrows denote the polarization contributions in the scattering plane. The circles denote the component $P_2$ that is perpendicular to the scattering plane; the radius of the circles represent the magnitude of $P_2$ and the dot (cross) denote the sign of $P_2$, positive (negative). SM predictions are in blue and CP violation predictions with $\xi=0.1\pi$ are in red. Note that the SM polarization always lies in the scattering plane ($P_2=0$), so the circles apply only to the CP violating case.
}
\label{fig:pol5}
\end{figure}

In summary of our findings, we show in Fig.\ \ref{fig:pol5} the polarization vector of Fig.\ \ref{fig:pol4} in the $W^+b\to th$ scattering plane. The solid blue arrows are the SM prediction, where all the arrows lie in the scattering plane, with length $P$. The predictions for $\xi=0.1\pi$ are shown by red symbols, where $(P_3,P_1)$ components are shown by arrows while $P_2$ components are given by circles as follows $|P_2|>0.7$ (large circles), $0.7>|P_2|>0.4$ (medium circles), $0.4>|P_2|>0.1$ (small circles) and $|P_2|<0.1$ (no circles); the signs of $P_2$ are denoted by the dots (positive) or crosses (negative) within the circles. In addition to the $W=400$ GeV results shown in Fig.\ \ref{fig:pol4}, we also give top polarizations for $W=$ 600 GeV~\cite{VBKHYZ}. The predicted pattern of the flow of top polarization in Fig.\ \ref{fig:pol5} will determine $\xi$ from the data, and thereby probe CP violation. The flow will also test the overall consistency of the model (our dimension-4 complex Higgs to top Yukawa coupling).
\\

\begin{acknowledgements}
 We are grateful to Junichi Kanzaki and Kentarou Mawatari for helpful discussions. YZ wishes to thank Tao Han and PITT PACC members for warm hospitality. This work has been supported in part by the U.S.\ Department of Energy under contract number DE-SC-0017647, and Grant-in-Aid for Scientific Research (No.\ 16F16321) from JSPS.
\end{acknowledgements}

\end{document}